\RequirePackage{fix-cm}
\documentclass[twocolumn,epjc3]{svjour3}  
\smartqed  
\RequirePackage{graphicx}
\RequirePackage{mathptmx}      
\RequirePackage[numbers,sort&compress]{natbib}
\RequirePackage[colorlinks,citecolor=blue,urlcolor=blue,linkcolor=blue]{hyperref}
\RequirePackage{siunitx}
\usepackage{orcidlink}
\usepackage{xcolor}
\definecolor{orcidlogocol}{HTML}{A6CE39}

\journalname{Eur. Phys. J. C}
\begin{document}

\title{The matrix optimum filter for low temperature detectors dead-time reduction
}


\author{Matteo Borghesi\thanksref{addr1,addr2} 
        \and
        Marco Faverzani\thanksref{addr1,addr2} 
        \and
        Cecilia Ferrari\thanksref{e1,addr3,addr4}\orcidlink{0000-0002-0838-2328} 
        \and
        Elena Ferri\thanksref{addr1,addr2} 
        \and
        Andrea Giachero\thanksref{addr1,addr2} 
        \and
        Angelo Nucciotti\thanksref{addr1,addr2} 
        \and
        Luca Origo\thanksref{addr1,addr2}
}

\thankstext{e1}{e-mail: cecilia.ferrari@gssi.it}


\institute{Dipartimento di Fisica “G. Occhialini”, Università di
	       Milano-Bicocca, Milan, 20126, Italy \label{addr1}
	       \and 
           INFN - Milano-Bicocca, Milan, 20126, Italy \label{addr2}
           \and
           Gran Sasso Science Institute (GSSI), L’Aquila, I-67100, Italy \label{addr3}
           \and
           INFN - Laboratori Nazionali del Gran Sasso, Assergi (L’Aquila), I-67100, Italy \label{addr4}
}

\date{Received: date / Accepted: date}

\maketitle

\begin{abstract}

Experiments aiming at high sensitivities usually demand for a very high statistics in order to reach more precise measurements. However, for those exploiting Low Temperature Detectors (LTDs), a high source activity may represent a drawback, if the events rate becomes comparable with the detector characteristic temporal response. Indeed, since commonly used optimum filtering approaches can only process LTDs signals well isolated in time, a non-negligible part of the recorded experimental data-set is discarded and hence constitute the \textit{dead-time}. In the presented study we demonstrate that, thanks to the matrix optimum filtering approach, the dead-time of an experiment exploiting LTDs can be strongly reduced.

\keywords{Low Temperature Detectors \and digital signal processing \and digital filters\and data processing methods \and dead-time reduction}
\end{abstract}

\section{Introduction}

Low Temperature Detectors (LTDs) are one of the most suitable detectors for experiments demanding a very high energy resolution for energy ranges up to \si{MeV}~\cite{Fiorini}. 
This request usually stems from the need of a high background rejection power and a good sensitivity for a low-probability fraction of the measured energy spectrum.
However, whenever the background or the signal rate becomes comparable to the LTDs characteristic temporal profile, the experiment may suffer from sensitivity loss. 
Indeed, in this case a non-negligible fraction of the collected data-set would be composed of pile-up on tail events, which have to be discarded. This is due to the fact that the optimum filter technique described in~\cite{Gatti} cannot correctly process pile-up on tail events, since it is non-causal. The amount of discarded data, due to the limited performances of this filter, constitutes the dead-time.

A large dead-time translates in a heavily reduced signal statistics that in turn worsen the target measurement. With the aim of reducing the amount of dead-time, we investigated an alternative approach, presented in~\cite{Fowler2015}, for the processing of LTD pulses.
In this paper we report the study carried out on this optimum filter~\cite{Fowler2015} (matrix filter in the following) in the framework of the HOLMES experiment~\cite{Alpert:2014lfa}. We probed the performances of this alternative optimum filtering technique on both real and simulated HOLMES microcalorimeters data by computing the experimental energy resolution and the projected dead-time.

\section{Pulse processing techniques}
\label{sec:filters}

In order to reduce noise contributions, microcalorimeters pulses are usually processed before estimating their interesting parameters. 
For example, in order to evaluate the energy of an event, a processing technique optimizing the evaluation of the pulse amplitude has to be exploited.
In this section we describe two pulse processing techniques for pulse amplitude evaluation.
The first one is the most known and widely used technique: the standard optimum filter of~\cite{Gatti}, (standard filter in the following).
The second one is the matrix optimum filter of~\cite{Fowler2015} (matrix filter in the following).
Both the two filtering techniques require that the processed pulses are acquired in record windows whose length is determined by the detector characteristic time profile in such a way that, within their record windows, pulses can fully recover.
Moreover, the two filters theoretical derivations assume that the pulses have a fixed shape and that the noise, to which the signals are subject, is ergodic.

\paragraph{Standard filter}

Let $S(\omega)$ be the Fourier Transform of the pulse ($S(t)$) and $N(\omega)$ the microcalorimeter noise power spectral density. The standard filter ($H$) is derived, as in~\cite{Gatti}, by maximizing the signal-to-noise ratio ($\rho$) of the processed pulse:
\begin{eqnarray}
	\rho^{2} \propto \frac{\left(\int_{-\infty}^{\infty}H(i\omega)S(\omega)e^{i\omega t_{\text{max}}}\text{d}\omega\right)^{2}}{\int_{-\infty}^{\infty}N(\omega)|H(i\omega)|^{2}\text{d}\omega}
	\label{eq:1}
\end{eqnarray}
where $t_{\text{max}}$ is the time at which the pulse reaches its maximum value.
Equation~\ref{eq:1} can be maximized by applying the Shwartz inequality, leading to:
\begin{equation}
	H(i\omega) = \frac{S^{*}(\omega)}{|N(\omega)|^{2}}e^{i\omega t_{max}}
\end{equation}
where $S^{*}(\omega)$ is the complex conjugate of the pulse Fourier Transform. The pulse amplitude is then estimated by evaluating the maximum of the filtered pulse ($f(t)$):
\begin{equation}
	f(t) = \int_{-\infty}^{\infty}H(i\omega)S(\omega)e^{i\omega t}\text{d}\omega
\end{equation}

\paragraph{Matrix filter}

The matrix filter is derived by maximizing the likelihood $L$ between the processed pulse and its model. Let ${\bf d}$ and ${\bf m}$ be respectively the pulse data vector and its model both of length $n$, corresponding to the number of pulse sampling points. 
Since the model generally depends on a certain number $n'$ of parameters, it is better to refer to the model as ${\bf m}({\bf p})$, where $\textbf{p}$ is the parameters vector of length $n'$.
If ${ R}$ is the data covariance noise matrix of dimensions $n\times n$, the likelihood between the model and the data results to be:
\begin{equation}
	L \propto \exp\left[-({\bf d} - {\bf m}(\textbf{p}))^{T}{R}^{-1}({\bf d} - {\bf m(\textbf{p})})\right]
	\label{eq:lik}
\end{equation}
Assuming that the microcalorimeters pulses always have the same shape and that their amplitude linearly depends on the energy released, it is possible to assume that the model {\bf m}(\textbf{p}) is equal to the average detector response ${\bf s}$ with length $n$ and unitary amplitude multiplied by just one parameter $p_{1}$: the pulse amplitude value. 
It is worth noting that this model would also work in general even if the relation between the pulse amplitude and the energy released is no more linear, provided that the conversion between these two parameters is well known.

In a more realistic picture, the pulses model should account for more elements, such as a constant vector modeling the eventual flat baseline over which the pulse rises. In this case the model {\bf m} can be written as:
\begin{equation}
{\bf m}(\textbf{p}) = p_{1}\times{\bf s} + p_{2}\times[1,\dots,1]
\end{equation}
where $p_{2}$ is the parameter representing baseline level. The model ${\bf m}(\textbf{p})$ can be therefore recast as $ {\bf m} = {\bf p}{M}$ where ${\bf p}$ is the parameters vector and $M$ the matrix collecting the pulse model elements of dimension $n\times n'$.

Under this assumption, the maximization of the likelihood in~\ref{eq:lik} returns the best estimation of the parameters vector, ${\bf \bar{p}}$, which results in:
\begin{equation}
{\bf \bar{p}} = \left(M^{T}R^{-1}M\right)\left(M^{T}R^{-1}\right){\bf d}
\end{equation}
From this relation it is possible to identify the filter $q$ as 
\begin{equation}
	{ q} = \left(M^{T}R^{-1}M\right)\left(M^{T}R^{-1}\right)
	\label{eq:filter}
\end{equation}
The filter $q$ accounts for both the processed pulse model components and  the noise covariance matrix as the standard filters do. Moreover, in order to get the pulse amplitude estimation, it is sufficient to compute the matrix inner product between the filter and the data.
Differently from the standard filter, the matrix one permits the introduction of additional elements to better model the processed data.
For example, an extra {\bf s} vector modeling a second pulse in the record would in principle allow the processing of pile-up on tail events.

We observed that this approach heavily depends on the temporal discretization of the recorded pulses.
Indeed, in order to mitigate any possible artifacts, e.g. non-correct amplitudes evaluation due to infra-sample pulses arrival times, in the matrix $M$ for every ${\bf s}$ array a second derivative vector of the averaged detector response is introduced.

\section{The HOLMES experiment data-set}

The HOLMES experiment, which aims to measure the electron anti-neutrino mass with a sensitivity of $\sim$\SI{2}{eV}, will exploit $1024$ Mo/Cu Transition-Edge Sensor (TES)~\cite{Irwin} based microcalorimeters implanted with $^{163}$Ho.
The HOLMES TES pulses are acquired with a derivative threshold trigger and recorded by means of windows with prefixed length, during which the trigger will be paralyzed.
Moreover, in order to also collect information on the pulses baseline, the record window is shifted backward in time, with respect to the pulse triggering time, of about one tenth of the window length. This part of the window is hence called \textit{pretrigger}.

\begin{figure}
	\centering
	\includegraphics[width=.9\linewidth]{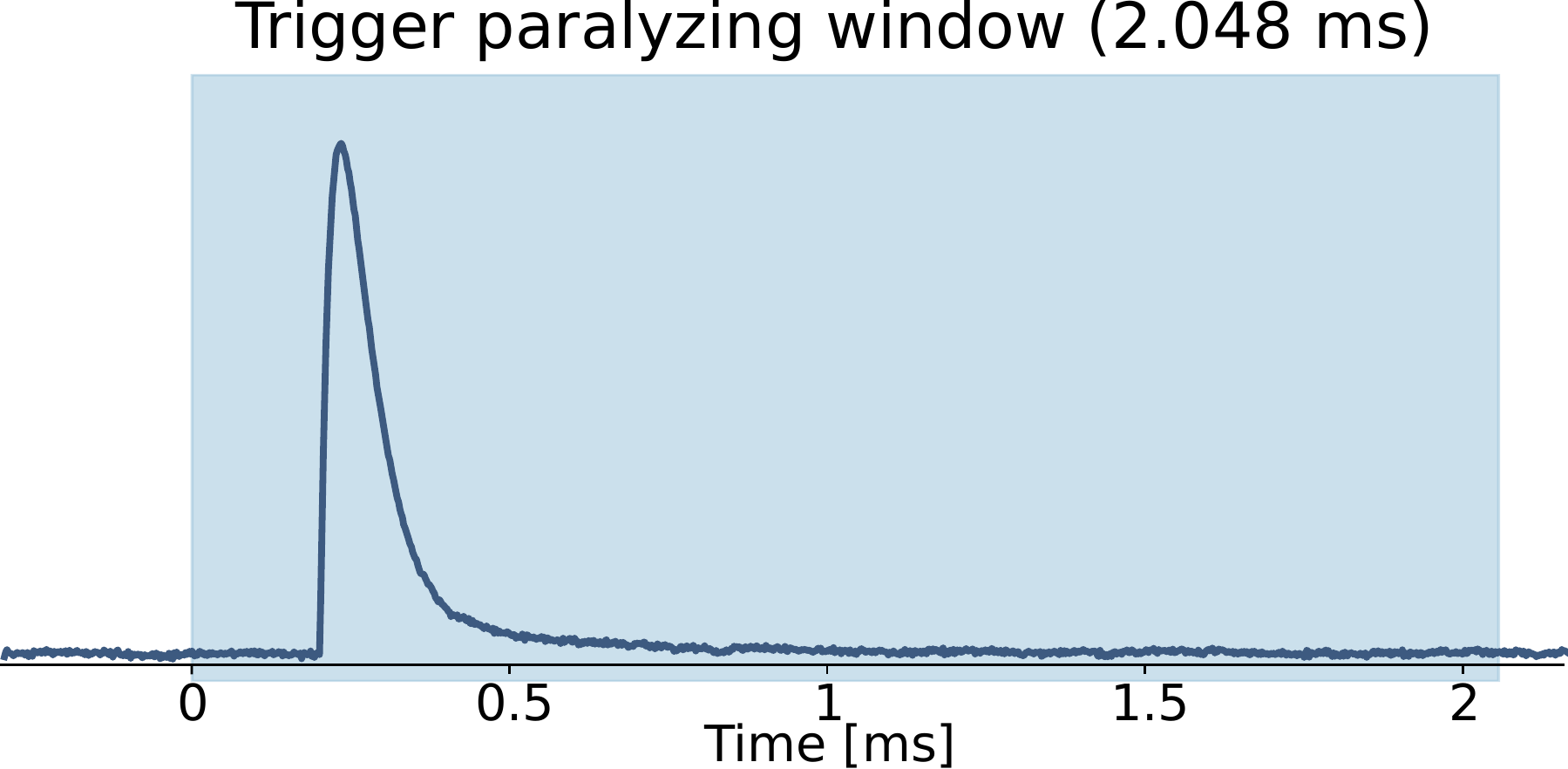}
	\caption{Example of a simulated pulse of a HOLMES microcalorimeter. In its \SI{2.048}{ms} record window (light-blue in the figure), the pulse is sampled with $1024$ points (out of which $100$ are of pretrigger) with a sampling frequency of \SI{500}{kHz}}
	\label{fig:0}       
\end{figure}

The window length is chosen on the basis of the microcalorimeters characteristic temporal response, in order to allow the full recovery of the signals to the baseline level within the window. 
As a rule of thumb, we suppose that a window length larger than five times the pulse decay time guarantees its complete recovery.
Assuming the temporal characteristics of the microcalorimeters studied in~\cite{Alpert2019}, a record window of \SI{2.048}{ms}, composed by $1024$ points (out of which $100$ are of pretrigger) with a sampling frequency of \SI{500}{kHz}, would be suitable for the experimental purposes.
Indeed with this choices the signals can fully recover within the window, as shown in figure~\ref{fig:0}, where a pulse simulated to reproduce those collected in the study presented in~\cite{Alpert2019} is shown. 

Each record window is then classified according to the features shown. If the pretrigger is characterized with a flat baseline and in the whole window only one pulse is present, it is tagged as a \textit{good} event. If the baseline is flat but the record features more than one pulse, it is classified as a \textit{multiple} event. If the baseline is non-flat, the record is labeled as \textit{bad-baseline}.

\section{Simulation of the data-set}

TESs are sensitive to temperature variations. If properly biased  with a voltage producing a current $I_{bias}$, their response to a release of energy E, corresponding to a power $P$, can be calculated by solving the thermal and electrical differential equation of the TES model~\cite{Irwin}. The so-computed response depends both on external parameters, such as the environment temperature ($T_{bath}$) and electrical circuit impedance ($L$), and on the TES properties, such as its resistance ($R_{TES}$), its temperature ($T$), its capacitance ($C$), the voltage ($V_{TES}$), the current flowing in it ($I_{TES}$) and the thermal conductance between the TES and the environment ($G$).

\begin{figure}
	\centering
	\includegraphics[width=.5\linewidth]{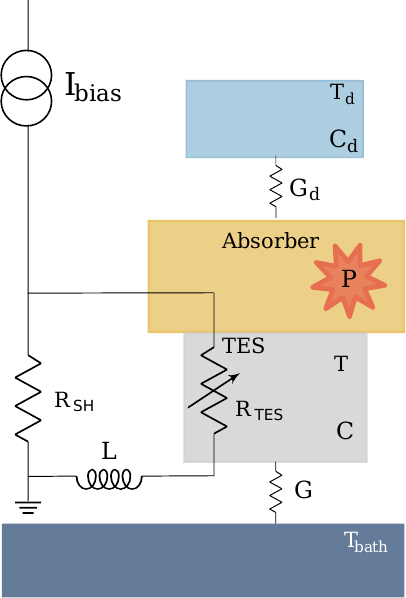}
	\caption{Two-body dangling model for HOLMES microcalorimeter pulses simulation. The TES is represented by the gray body, the absorber part of the microcalorimeter where the $^{163}$Ho is implanted is depicted in yellow and the dangling body is illustrated in light-blue. Whenever a $^{163}$Ho decay occurs in the absorber, a release of energy E, corresponding to a power P (depicted in red in the scheme), triggers the microcalorimeter response which depends both on the electric and thermal circuit parameters.}
	\label{fig:2}       
\end{figure}

For the HOLMES TESs a good agreement was found between real pulses and those simulated with the two-body dangling model~\cite{Borghesi, Maasilta} presented in figure~\ref{fig:2} and described by the following set of equations:
\begin{equation}
	\small
	\left\{\begin{array}{l}
		C \frac{\text{d}T_{TES}}{\text{d}t} = -K(T^{n}-T_{bath}^{n}) -K_{d}(T^{n_{d}}-T_{d}^{n_{d}}) -I_{TES}^{2}R_{TES} +P \\
		\\
		C_{d} \frac{\text{d}T_{d}}{\text{d}t} =  -K_{d}(T^{n_{d}}-T_{d}^{n_{d}}) \\
		\\
		L \frac{\text{d}I}{\text{d}t} = V_{TES} -I_{TES}R_{L} -I_{TES}R_{TES} 
	\end{array}\right.
	\label{eq:TESeqdangling}
\end{equation}
where, $C_{d}$ and $T_{d}$ are the heat capacity and the temperature of the dangling body in the model (depicted in light-blue in figure~\ref{fig:2}) and  $K$ and $K_{d}$ are constants depending on the thermal conductances $G$ and $G_{d}$, respectively.

Equations in~\ref{eq:TESeqdangling} can be solved with a fourth order Runge-Kutta method, by discretizing the time variable as to simulate the HOLMES data acquisition system. In order to study realistic data and simulate the temporal jitter effect, a pulse interpolation with a random shift within the interval $[-0.5,0.5]$ in terms of sampling points is applied.
The so-obtained pulses were then superimposed to random samples of noise.
These latter were obtained by extracting the parameters of the AutoRegressive Moving Average model (ARMA(p,q))~\cite{Box} from the comparison between theoretical and experimental noise power spectral densities of the device.

\begin{figure}
	\includegraphics[width=\linewidth]{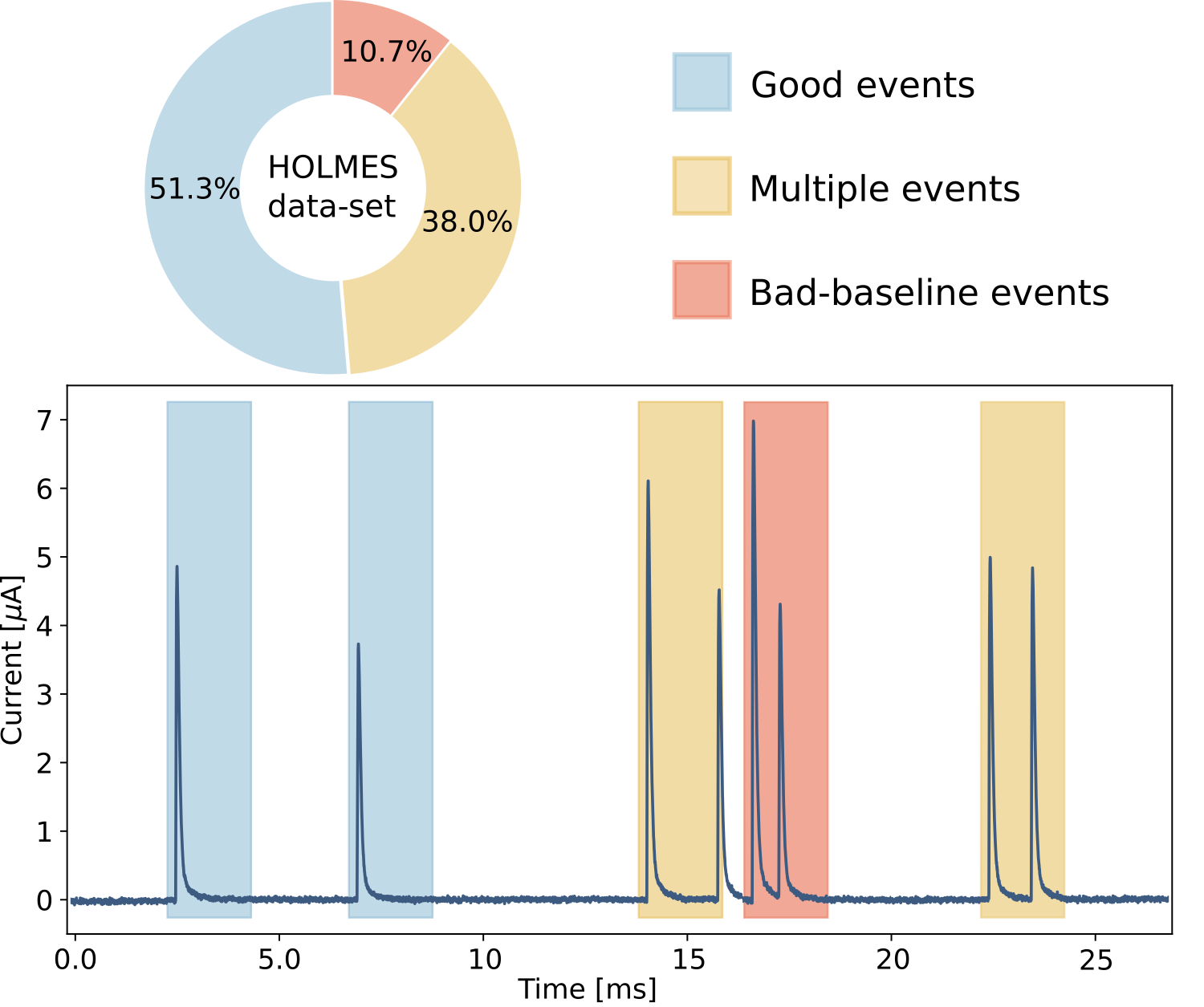}
	\caption{Results of a HOLMES experiment TES simulation: composition of the HOLMES experiment data-set (top); part of raw data flow from a HOLMES microcalorimeter with triggered record windows colored as the classification of the events (bottom).}
	\label{fig:1}       
\end{figure}

\begin{figure}
	\centering
	\includegraphics[width=\linewidth]{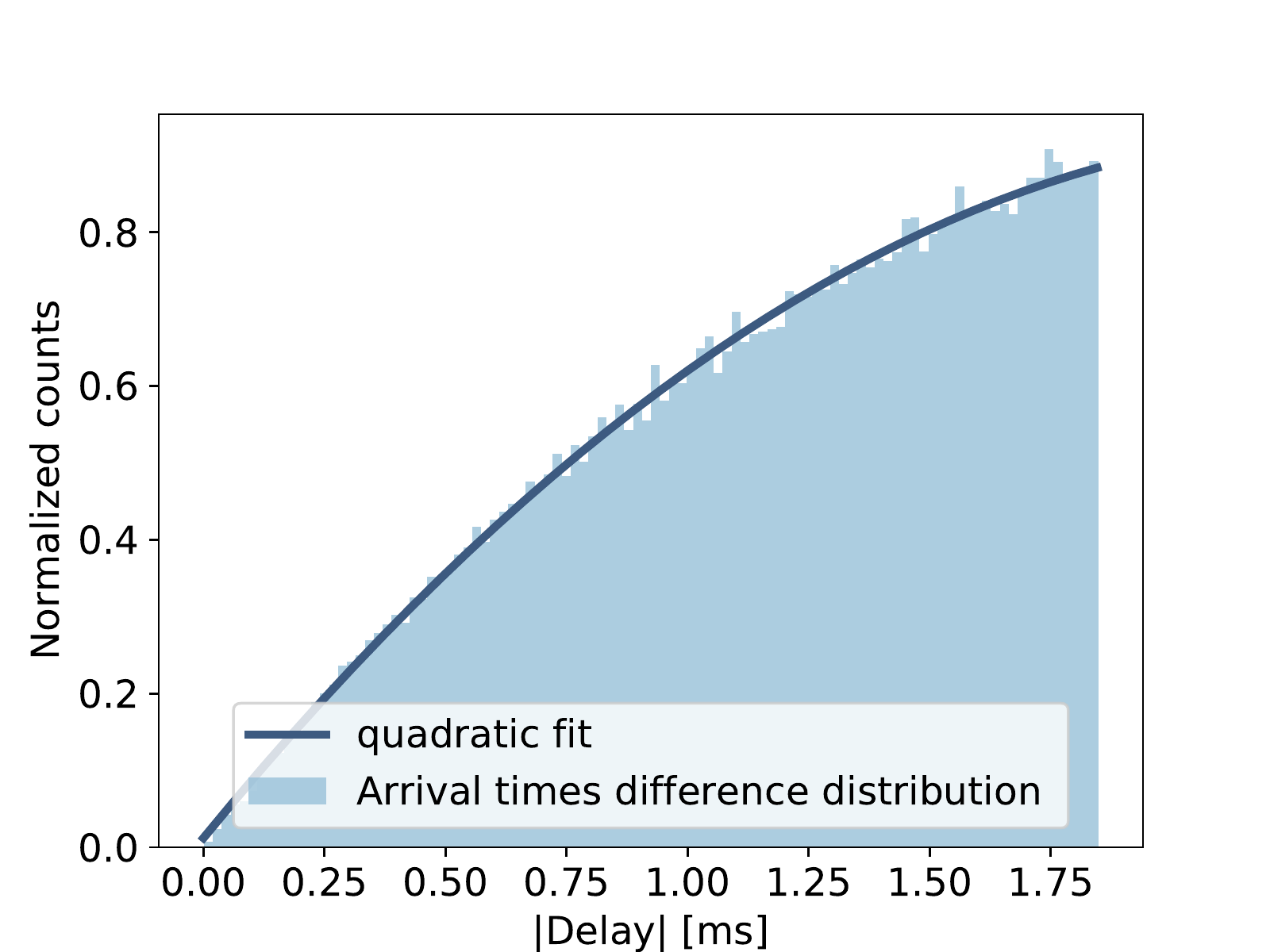}
	\caption{Distribution of arrival times difference (delay) between the triggered pulse and its previous one in {\it bad-baseline} events. It follows a quadratic trend.}
	\label{fig:3}       
\end{figure}

In order to properly simulate the data-set to analyze with the two different filters, a Monte Carlo study considering the HOLMES experiment final configuration ($\sim1000$ detectors with \SI{300}{Bq} of $^{163}$Ho activity each~\cite{Alpert:2014lfa}), was performed. 
With a conservative approach, in this study, we labeled as {\it bad-baseline} every record window whose starting point is less than \SI{1.848}{ms} away the previous pulse arrival time.
The results of this simulation are reported in figure~\ref{fig:1}. The study reveals that for a \SI{300}{Bq} of $^{163}$Ho source, which is the target activity per single HOLMES microcalorimeter~\cite{Alpert:2014lfa}, only half of the recorded events are \textit{good}, and hence processable with the standard filter.
Moreover, thanks to the same study, the distribution of arrival times differences between {\it bad-baseline} events triggered pulses and their previous ones is estimated. This outcome, which is perfectly fit with a quadratic function, is reported in figure~\ref{fig:3}. For sake of clarity, from now on, we will simply refer to the difference between two consecutive pulses arrival times as \textit{delay}.

For this work we have simulated a HOLMES data-set following the results of the Monte Carlo of figure~\ref{fig:1} .
For sake of simplicity, the data-set we simulated accounts for \textit{good} events, \textit{multiple} events with two pulses only and {\it bad-baseline} events featuring one pulse, with the correct proportions and temporal distributions. \textit{Good} and {\it bad-baseline} events were generated by assuming a release of \SI{3000}{\eV}. The choice of this energy value is dictated by the fact that it is very close to the $^{163}$Ho spectrum end-point (at $2833$ \SI[parse-numbers=false]{\pm30(stat)}{} \SI[parse-numbers=false]{\pm 15(sys)}{eV}~\cite{Eliseev}), which is the HOLMES experiment region of interest (ROI) and therefore we expect to obtain, for this energy value, the same energy response that would be observed in the ROI. 
Moreover, \textit{multiple} events were split into two categories: \textit{Multi1} with first pulses corresponding to \SI{3000}{\eV} and second pulses randomly generated according to the $^{163}$Ho spectrum and \textit{Multi2} representing the \textit{viceversa}.

\section{Detector responses with the two different filters}

In order to estimate the amount of expected dead-time depending on the exploited filter, it is necessary to study how the detector response ($R(E)$), which is the distribution of measured energies in response to a monochromatic input, is modified accordingly to the considered sub-set of events.
For a microcalorimeter, assuming ergodic noise, $R(E)$ is a Gaussian distribution when only {\it good} events are analyzed. The standard deviation of this distribution is defined by the filter performances in discriminating the signal to noise contributions. 
We verified with the study reported in~\cite{Ferrari} that the two filters performances are comparable in terms of energy resolution measured as the full width at half maximum (FWHM) of the K$_{\alpha 2}$ X-ray peak of $^{55}$Mn. Indeed we found for the matrix filter an energy resolution of \SI{4.61\pm0.14}{eV} at \SI{5.89875}{\keV} in line with the one measured with the standard filter, of \SI{4.5\pm0.1}{eV}~\cite{Alpert2019} at the same energy peak, obtained with the same set of data.

As one would expect, the detector response is altered once non {\it good} events are processed, since these records deviate from the model one ($S(t)$ for the standard filter and {\bf m(p)} for the matrix one). 
Indeed, in this case, the processing of non {\it good} pulses alters $R(E)$ in such a way that is not possible to find a correct analytical model.
Therefore, it is better to reduce the set of analyzed data up to a point in which the deviations from the analytical $R(E)$ (Gaussian in our case) no more introduce systematics in the measured parameters of interest.
To perform this kind of study, we analyzed the different detector responses obtained with the two filters by varying the set of analyzed data.

\subsection{Standard filter detector responses}

\begin{figure}
	\includegraphics[width=\linewidth]{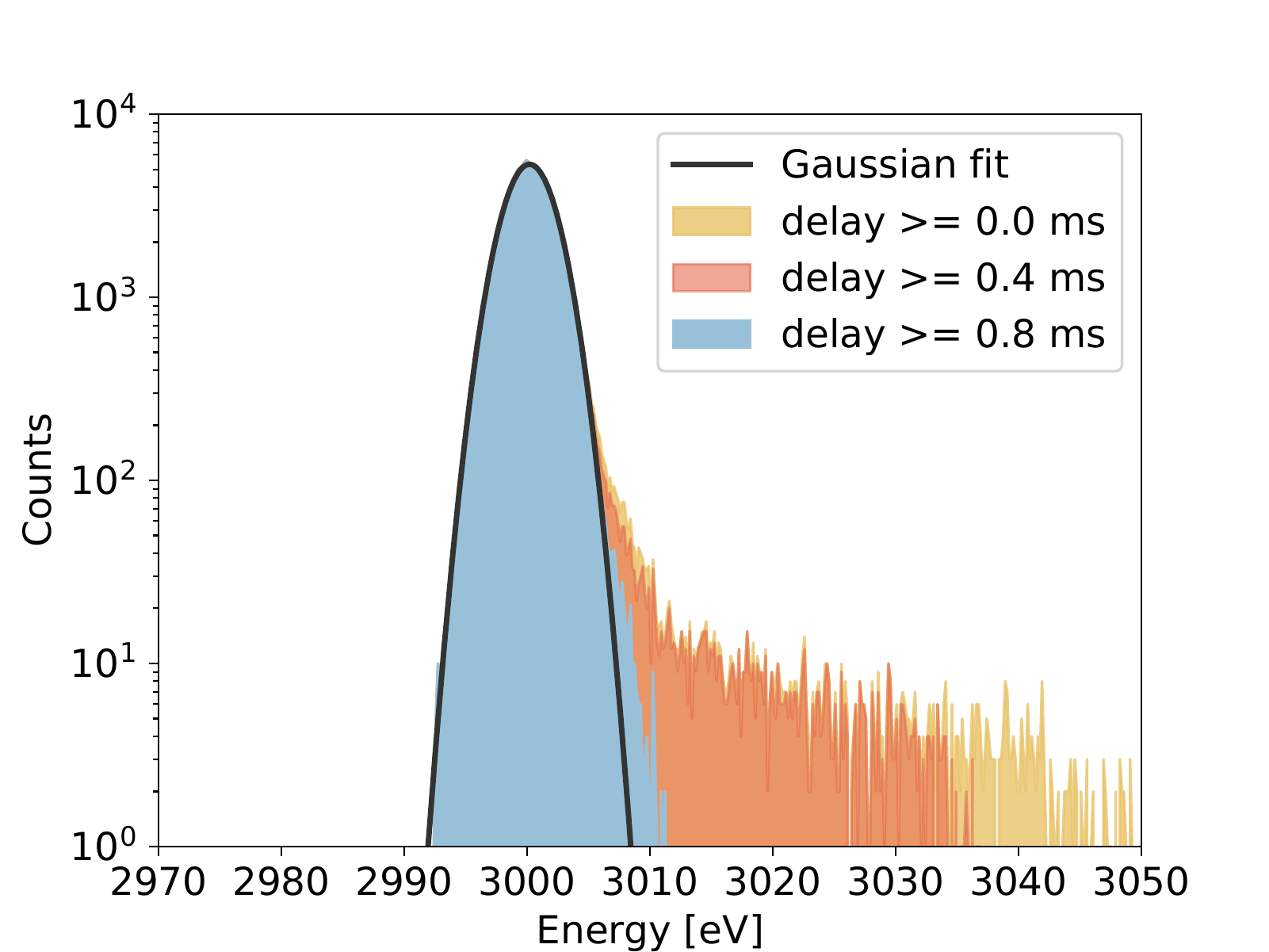}
	\caption{Detector responses ($R(E)$) generated with the application of the standard filter to different sub-sets of the data-set. In yellow $R(E)$ obtained with the complete data-set (\textit{good} plus all single {\it bad-baseline} events), in red $R(E)$ including only {\it bad-baseline} single pulses with delay greater than \SI{0.4}{\ms} and in light-blue $R(E)$ considering those with delay greater than \SI{0.8}{\ms}. The plot also shows the Gaussian fit (in black) of the yellow detector response. The values of the Gaussian parameters from the fit are: mean$=3000.191\pm$\SI{0.005}{\eV}, standard deviation$=1.993\pm$\SI{0.004}{\eV} and norm$=(2.658\pm0.007)\times10^{4}$.}
	\label{fig:4}       
\end{figure}

Since the standard filter, being non-causal, cannot process events with more than one pulse, we can only try to include in the processing also single {\it bad-baseline} records, constituting only the \SI{6.1}{\percent} of the data-set.
The detector responses are obtained by considering several sub-set of data created by defining different lower thresholds on the delay, varying with a step of \SI{0.1}{\ms}.
Accordingly to the studied sub-set of data, the standard filter detector response is modified. In particular, the integration of more single {\it bad-baseline} pulses in the analyzed data-set produce a bigger tail at higher energies, leaving unaltered the Gaussian distribution peaked at \SI{3000}{\eV}.
As a proof for this, three example of detector responses characterized by three different lower thresholds in the delay are depicted in figure~\ref{fig:4}.

\subsection{Matrix filter detector responses}

The matrix filter can in principle be applied to each class of events. 
However, even if in case of {\it good} events only its performances in terms of computational cost are similar to the standard filter, in case of non {\it good} events the matrix filter has to be adjusted case-by-case in order to correctly process each pulse in the records. This implies that this filter, for the processing of non {\it good} events, has to be applied offline. 

In order to construct the matrix filter detector responses, as in the case of the standard filter, different sub-sets of data are created.
In this analysis we considered \textit{good} pulses, {\it bad-baseline} signals, $Multi1$ records and $Multi2$ events with delays smaller than \SI{1.756}{ms}. The latter limit is a consequence of the fact that, with longer delays, the pulse cannot merely reach its maximum in the record window (which lasts for just \SI{1.848}{ms} after the first pulse arrival time). The value for this limit was set by requiring that the detector response $R(E)$, at large lower thresholds on the delay, is well-fitted by a Gaussian distribution (p-value of fit $\sim1$).

In both cases, {\it bad-baseline} and {\it multiple} events, before computing the matrix filter $q$, the information on the delay (temporal difference between the record pulse and the prior one, in the {\it bad-baseline} case, and the temporal difference between the two pulses in the record, in the {\it multiple} event case) has to be extracted from the records.
To assess pulses arrival times, we exploit a simple threshold triggering technique on the pulses rising edge. 
Notice that, in the case of {\it bad-baseline} records, to evaluate the delay we need to have the information on the event recorded before the analyzed one. This is always possible in the HOLMES experiment, since  an absolute temporal value referring to the start of the data taking (\textit{timestamp}) is saved for each recorded event.
Once the delay is assessed, two extra vectors are included in the matrix model $M$ defined in section~\ref{sec:filters}: one average detector response \textbf{s} shifted in time by a factor equal to the computed delay and its second derivative vector. By shifting forward (backward) in time the \textbf{s} vector, for events classified as {\it multiple} ({\it bad-baseline}), some void points at the beginning (end) of the vector are created. These void points are filled with zeros in both cases of {\it multiple} and {\it bad-baseline} events.  
On the one hand, this represents the most natural choice for the {\it multiple} cases, in which the pulse is shifted forward and it does not contribute at all to the model prior to its arrival.
On the other hand, this choice, for the {\it bad-baseline} case, was dictated by the results of an analysis revealing that, by completing the shifted \textbf{s} vector with the average value of its last $10$ points, the detector response showed a slightly worse resolution.

\begin{figure}
	\includegraphics[width=\linewidth]{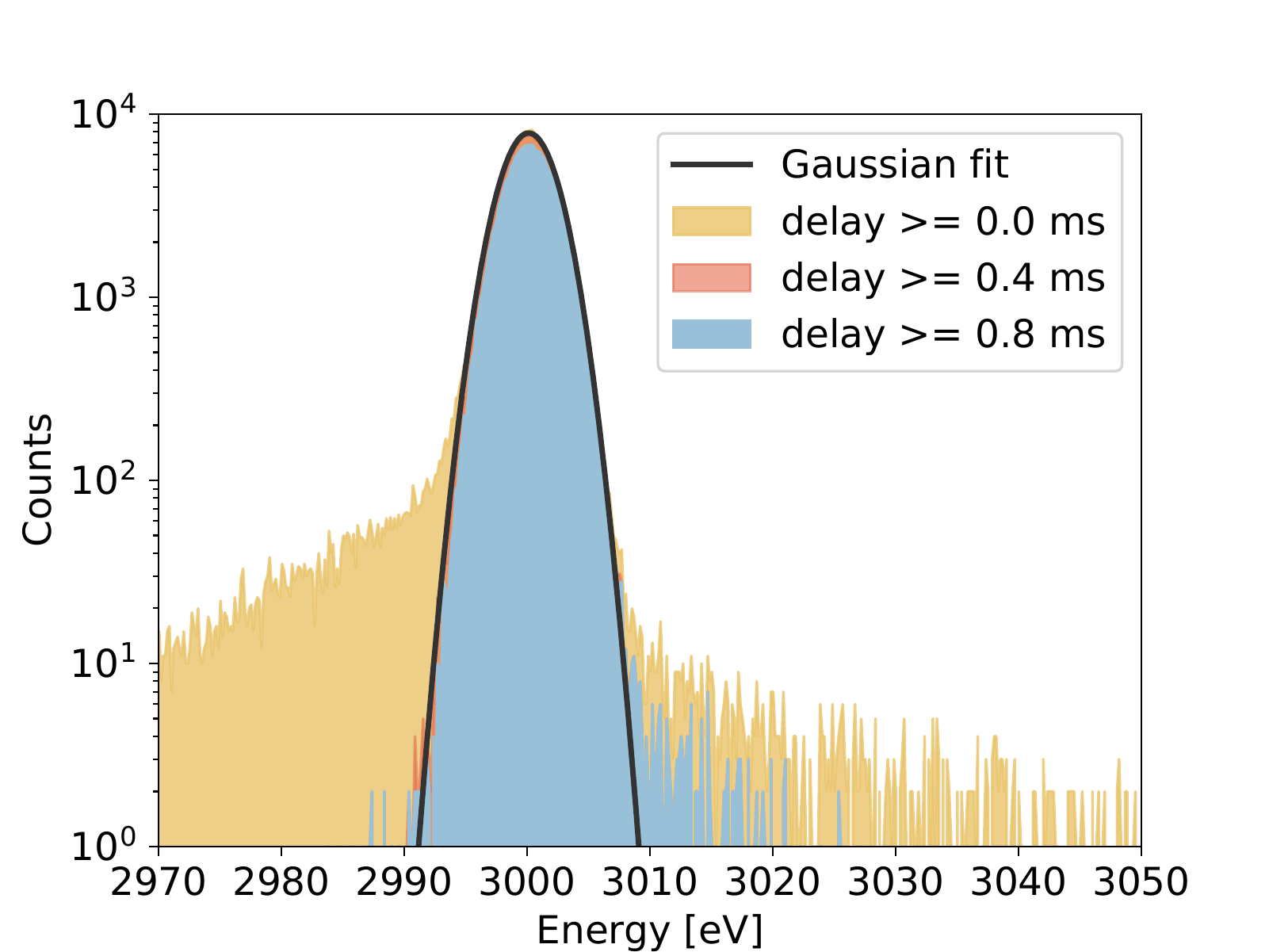}
	\caption{Detector responses ($R(E)$) generated with the application of the matrix filter to different sub-sets of the data-set, by omitting the evaluation of $Multi2$ events with a delay larger than \SI{1.756}{\ms}. In yellow $R(E)$ obtained with the complete data-set (\textit{good} plus all {\it bad-baseline} and all {\it multiple} events), in red $R(E)$ including only {\it bad-baseline} and {\it multiple} events with delay greater than \SI{0.4}{\ms} and in light-blue $R(E)$ considering those with delay greater than \SI{0.8}{\ms}. The plot also shows the Gaussian fit (in black) of the yellow detector response. The values of the Gaussian parameters from the fit are: mean$=3000.132\pm$\SI{0.004}{\eV}, standard deviation$=2.124\pm$\SI{0.003}{\eV} and norm$=(4.196\pm0.008)\times10^{4}$.}
	\label{fig:5}       
\end{figure}

In figure~\ref{fig:5} three examples of matrix filter detector responses characterized by three different lower thresholds on the delay (applied both to {\it multiple} and {\it bad-baseline} events) are reported. Differently from the standard filter case, one can notice that, by increasing the lower threshold on the delay, not only the distribution tails are modified, but also the Gaussian peak.
Indeed, as clearly visible from the comparison of the red and light-blue $R(E)$ of figure~\ref{fig:5}, a significant portion of counts at the Gaussian peak centered at \SI{3000}{\eV} are removed. This means that, differently from the standard filter case, this filter correctly estimate also the amplitudes of the pulses belonging to events with small delays.
However, we have observed that there exists a lower threshold on the delay below which it seems that the amplitudes are no more correctly reconstructed causing a broadening of the detector response (as one can evince from the yellow distribution in figure~\ref{fig:5}).
We found out that this effect cannot be ascribed to the matrix filter but rather to the intrinsic non-linearity of the TES response to multiple releases of energy very close in time, introduced by the equations in~\ref{eq:TESeqdangling}.   
This was proved by an analysis performed on \textit{multiple} events generated by linearly superimposing simulated single pulses. Indeed, for this latter case we obtained a way less broaden detector response.

\section{Dead-time evaluation for standard and matrix filters}

In order to evaluate the expected experimental dead-time according to the exploited filter, we studied the induced systematic effects, caused by the deviation of the detector response from the analytical one (Gaussian distribution), on the neutrino mass, which is the parameter the HOLMES experiment aims to assess by means of a fit of the $^{163}$Ho measured spectrum.
With the detector responses previously obtained, we convoluted, bin-by-bin, $10$ simulated $^{163}$Ho spectra comprising $10^{9}$ events, which correspond to the hypothetical amount of energy releases expected in one TES implanted with a \SI{300}{Bq} activity source in four months of data-taking. We then fitted the obtained convoluted histograms with a $^{163}$Ho spectrum profile smeared with a Gaussian distribution with fixed variance (obtained by the Gaussian fit of the responses $R(E)$). In the fit procedure the squared neutrino mass, the spectrum normalization and its end-point were left as free fit parameters. For this study we decided to set \SI{0.1}{\eV^{2}} as the preliminary maximum acceptable absolute value of the systematic effect on $m_{\nu}$. This value will be correctly tuned once the final configuration of the experiment will be defined. 

\begin{figure}
	\includegraphics[width=\linewidth]{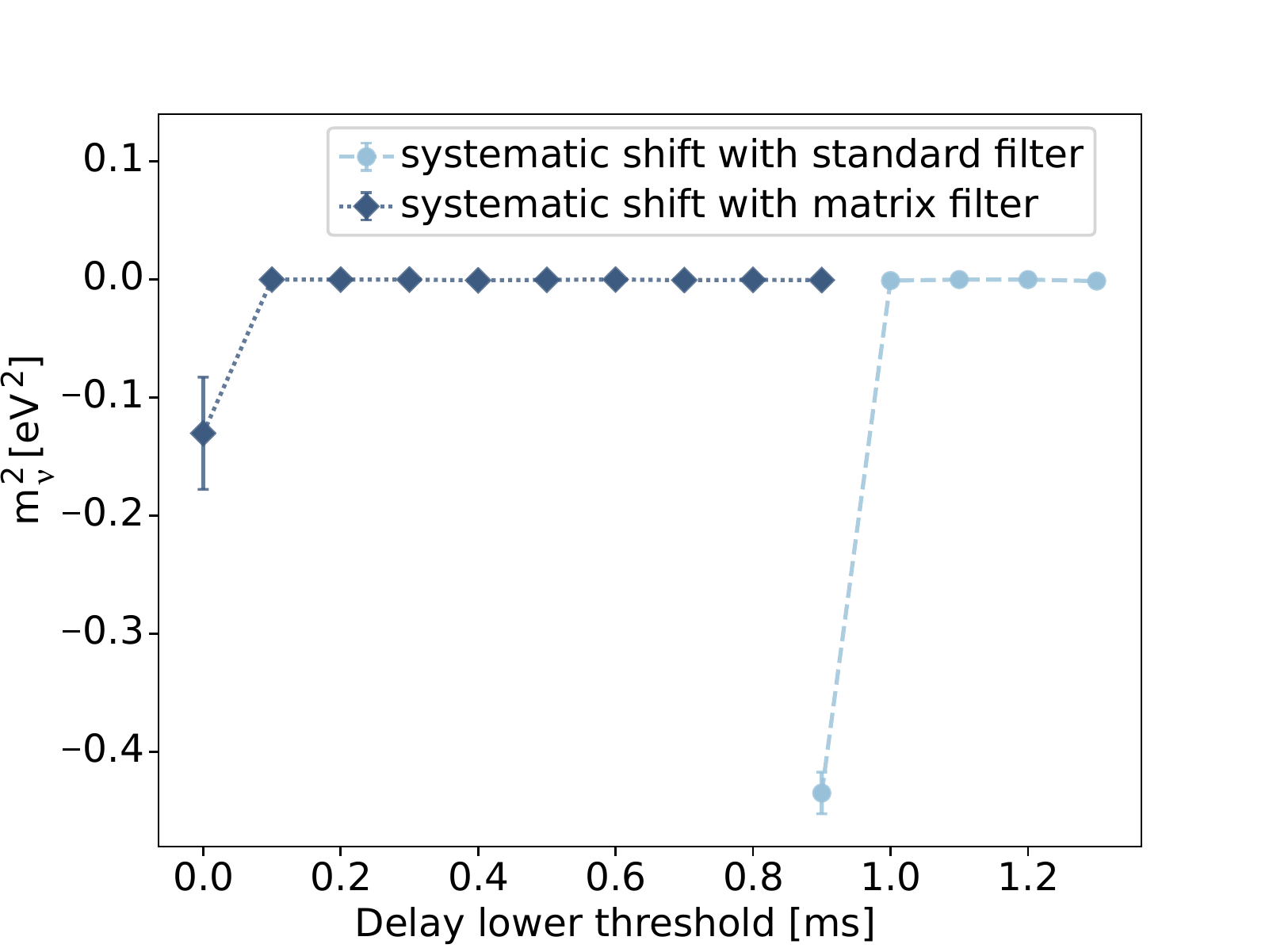}
	\caption{Systematic $m_{\nu}^{2}$ shift introduced by ignoring the non Gaussian detector response $R(E)$ (computed for several lower thresholds on the delay with a step of \SI{0.1}{\ms}) caused by the application of both standard (light-blue) and matrix (blue) filters. In this study, in a conservative way, \SI{0.1}{\eV^{2}} was established as maximum acceptable absolute value of the systematic shift on $|m_{\nu}^{2}|$.}
	\label{fig:6}       
\end{figure}

In figure~\ref{fig:6}, the average of the resulting values for the squared neutrino mass is plotted for each lower threshold on the delay, we have considered in the study.
As we can clearly see from figure~\ref{fig:6}, the standard filter already produces a detector response that induce a $|m_{\nu}^{2}|$ shift larger than \SI{0.1}{\eV^{2}} for a lower threshold on the delay of between \SI{0.9}{ms} and \SI{1.0}{ms}, while for the matrix filter this happens between \SI{0.0}{ms} and \SI{0.1}{ms}, due to the non-linearity effects. 

\begin{figure}
	\centering
	\includegraphics[width=\linewidth]{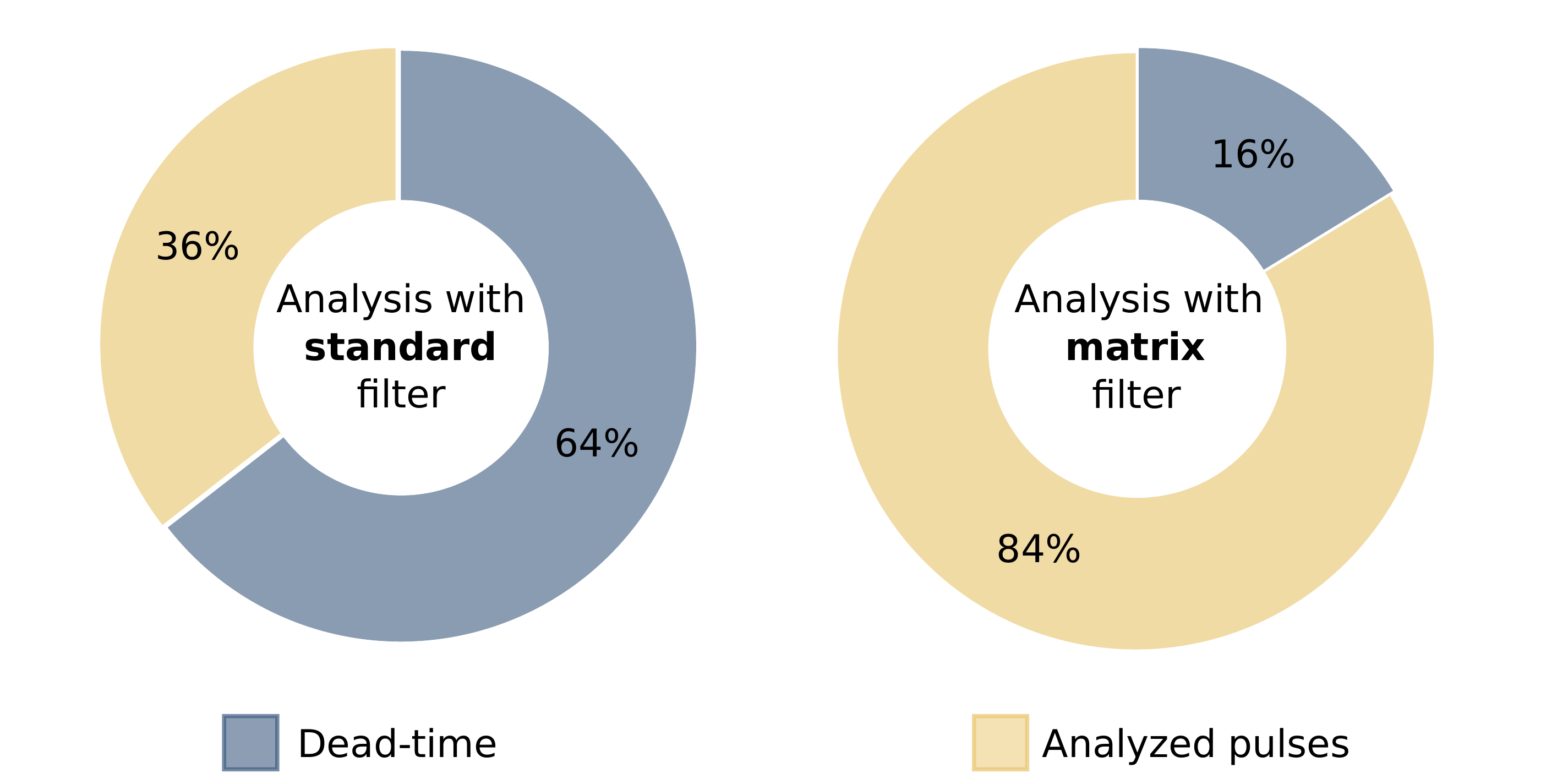}
	\caption{Dead-time (fraction of pulses that have to be discarded in the analysis phase) for both standard and matrix filters. Percentages are computed on a pulse-based and not event-based analysis.}
	\label{fig:7}       
\end{figure}

Once the lower threshold on the delay is defined, set to \SI{1.0}{\ms} for the standard filter and to \SI{0.1}{\ms} for the matrix one, we can compute the fraction of the data-set that can be processed according to the considered filter.
In a conservative way, we decided, in the computation of the expected dead-time, to discard a priori the fraction of the data-set composed by \textit{multiple} and {\it bad-baseline} events accounting for more than two pulses.
The results of this study, expressed in terms of percentages of processable pulses, are reported in figure~\ref{fig:7}, both for the standard and matrix filters.
We can conclude that in terms of expected dead-time, the matrix filter represents a great advantage for the HOLMES experiment. Indeed, it reduces the dead-time by a factor of about four.

\section{Conclusion}

The matrix filter of~\cite{Fowler2015} for microcalorimeters pulses processing represents a great alternative to the standard optimum filter, described in~\cite{Gatti}. Indeed, it provides very similar energy resolutions and, at the same time, it also allows the analysis of non {\it good} events. 
This property is of great advantage for experiments exploiting microcalorimeters subject to very high source activities. Indeed, as demonstrated with this study in the specific framework of the HOLMES experiment, it can strongly reduce the expected dead-time. The obtained reduction of dead-time, which translates into an increased statistics for the measurement of the neutrino mass of a factor greater than two, allows to improve the experimental sensitivity  on $m_{\nu}$. The results found in this study show that by exploiting the matrix filter instead of the standard one for the HOLMES experiment, the neutrino mass sensitivity, which scales with the inverse of the fourth root of the statistics, improves 
of about \SI{19}{\percent} .

\begin{acknowledgements}
This work was supported by the European Research Council (FP7/2007-2013), under Grant Agreement HOLMES no.340321, and by the INFN Astroparticle Physics Commission 2 (CSN2). We also acknowledge the support from the NIST Innovations in Measurement Science program for the TES detector development.
\end{acknowledgements}





\end{document}